\documentstyle [12pt,eqsecnum,aps] {revtex}
\tighten
\draft
\input epsf
\newcommand\fverb{\setbox\pippobox=\hbox\bgroup\verb}
\newcommand\fverbdo{\egroup\medskip\noindent%
			\fbox{\unhbox\pippobox}\ }
\newcommand\fverbit{\egroup\item[\fbox{\unhbox\pippobox}]}
\newbox\pippobox

\newcommand{\gtwid}{\mathrel{\raise.3ex\hbox{$>$\kern-.75em\lower1ex
\hbox{$\sim$}}}}
\newcommand{\ltwid}{\mathrel{\raise.3ex\hbox{$<$\kern-.75em\lower1ex
\hbox{$\sim$}}}}
\newcommand{\be}{\begin{equation}}
\newcommand{\ee}{\end{equation}}
\newcommand{\ba}{\begin{eqnarray}}
\newcommand{\ea}{\end{eqnarray}}

\newcommand{\beq}{\begin{equation}}
\newcommand{\eeq}{\end{equation}}
\newcommand{\beqs}{\begin{eqnarray}}
\newcommand{\eeqs}{\end{eqnarray}}
\newcommand{\lsim}{\mathrel{\raisebox{-.6ex}{$\stackrel{\textstyle<}{\sim}$}}}

\begin{document} 
\begin{titlepage}

\begin{center}

{ \LARGE Neutrino Oscillations with Two $\Delta m^2$ Scales} 

\vspace{1.1cm}

{\large Irina Mocioiu${}$\footnote{email: 
mocioiu@insti.physics.sunysb.edu } 
and Robert Shrock${}$\footnote{email: 
shrock@insti.physics.sunysb.edu }}\\
\vspace{18pt}
 C.N. Yang Institute for Theoretical Physics \\
 State University of New York \\
 Stony Brook, NY 11794-3840 \\
\end{center}
\vskip 0.6 cm

\begin{abstract}
\vspace{2cm}

An approximation that is often used in fits to reactor and
atmospheric neutrino data and in some studies of future neutrino oscillation
experiments is to assume one dominant scale, $\Delta m^2$, of neutrino mass
squared differences, in particular, $\Delta m^2_{atm} \sim 3 \times 10^{-3}$
eV$^2$.  Here we investigate the corrections to this approximation arising from
the quantity $\Delta m^2_{sol}$ relevant for solar neutrino oscillations,
assuming the large mixing angle solution.  We show that for values of
$\sin^2(2\theta_{13}) \sim 10^{-2}$ (in the range of interest for long-baseline
neutrino oscillation experiments with either intense conventional neutrino
beams such as JHF-SuperK or a possible future neutrino factory) and for $\Delta
m^2_{sol} \sim 10^{-4}$ eV$^2$, the contributions to $\nu_\mu \to\nu_e$
oscillations from both CP-conserving and CP-violating terms involving
$\sin^2(\Delta m^2_{sol}L/(4E))$ can be comparable to the terms involving
$\sin^2(\Delta m^2_{atm}L/(4E))$ retained in the one-$\Delta m^2$
approximation.  Accordingly, we emphasize the importance of performing a full
three-flavor, two-$\Delta m^2$ analysis of the data on $\nu_\mu \to \nu_e$
oscillations in a conventional-beam experiment and $\nu_e \to \nu_\mu$,
$\bar\nu_e \to \bar\nu_\mu$ oscillations at a neutrino factory.  We also
discuss a generalized analysis method for the KamLAND reactor experiment, and
note how the information from this experiment can be used to facilitate the
analysis of the subsequent data on $\nu_\mu \to \nu_e$ oscillations.  Finally,
we consider the analysis of atmospheric neutrino data and present calculations
of matter effects in a three-flavor, two-$\Delta m^2$ framework relevant to
this data and to neutrino factory measurements.  

\end{abstract}

\end{titlepage}

\section{Introduction}

There is increasingly strong evidence for neutrino oscillations, and thus
neutrino masses and lepton mixing.  All solar neutrino experiments that have
reported results (Homestake, Kamiokande, SuperKamiokande, SAGE and GALLEX/GNO,
SNO) show a significant deficit in the neutrino fluxes coming from the Sun
\cite{sol}.  This deficit can be explained by oscillations of the
$\nu_e$'s into other weak eigenstate(s).  The currently favored region of
parameters to fit this data is the solution characterized by a neutrino mass
squared difference $2 \times 10^{-5} \lsim \Delta m^2_{sol} \lsim 2 \times
10^{-4}$ eV$^2$ and a large mixing angle (LMA), $\tan^2 \theta_{12} \sim 0.4$
\cite{sol}-\cite{othersol}.  Solutions yielding lower-likelihood fits to the
data include the small-mixing angle (SMA) solution, with strong
Mikheev-Smirnov-Wolfenstein (MSW) matter enhancement \cite{msw}, and $\Delta
m^2_{sol} \sim 0.5 \times 10^{-5}$ eV$^2$, $\tan^2 \theta_{12} \sim 4\times
10^{-4}$, the LOW solution with $\Delta m^2_{sol} \sim 10^{-7}$ eV$^2$ and
essentially maximal mixing.

Another piece of evidence for neutrino oscillations is the atmospheric neutrino
anomaly, observed by Kamiokande \cite{kam}, IMB \cite{imb}, Soudan
\cite{soudan2}, SuperKamiokande (also denoted SuperK, SK) with the highest
statistics \cite{skatm}, and MACRO \cite{macro}.  The SuperK experiment has fit
its data by the hypothesis of $\nu_{\mu} \to \nu_\tau$ oscillations with
$\Delta m^2_{atm} \sim 3 \times 10^{-3}$ eV$ ^2$ and maximal mixing, $\sin^2 2
\theta_{atm} = 1$.  The possibility of $\nu_\mu \to \nu_s$ oscillations
involving light electroweak-singlet (``sterile'') neutrinos has been disfavored
by SuperK, and the possibility that $\nu_\mu \to \nu_e$ oscillations might play
a dominant role in the atmospheric neutrino data has been excluded both by
SuperK and, for the above value of $\Delta m^2_{atm}$, by the Chooz and Palo
Verde reactor antineutrino experiments \cite{chooz,paloverde}. The K2K
long-baseline neutrino experiment between KEK and Kamioka has also reported
results \cite{k2k} which are consistent with the SuperK fit to its atmospheric
neutrino data.  The LSND experiment has reported evidence for $\bar\nu_\mu \to
\bar \nu_e$ and $\nu_{\mu} \to \nu_e$ oscillations with $\Delta m^2_{LSND} \sim
0.1 - 1$ eV$^2$ and a range of possible mixing angles \cite{lsnd,lsndn}. This
result is not confirmed, but also not completely ruled out, by a similar
experiment, KARMEN \cite{karmen}.  The solar and atmospheric data can be fit in
the context of three-flavor neutrino mixing, and we shall work within this
context. 

The fact that these inferred values of neutrino mass squared differences
satisfy the hierarchy $|\Delta m^2_{sol}| << |\Delta m^2_{atm}|$ has led to a
commonly used approximation in fits to the reactor and atmospheric neutrino
data, and in studies of CP-conserving effects in terrestrial neutrino
oscillation experiments.  In this approximation, which we shall denote the
one-$\Delta m^2$ approximation (1DA), one neglects $\Delta m^2_{sol}$ compared
with $\Delta m^2_{atm}$.  For certain neutrino oscillation transitions, such as
$\nu_\mu \to \nu_\tau$, this is an excellent approximation.  It is worthwhile,
however, to have a quantitative evaluation of the corrections to this
approximation and a determination of the ranges of parameters where these
corrections could become significant.  Indeed, for sufficiently small values of
the lepton mixing angle $\theta_{13}$ (defined below in eq. (\ref{u})), e.g.,
$\sin^2 2\theta_{13} \sim 10^{-2}$, and sufficiently large values of $\Delta
m^2_{sol}$, e.g., $\Delta m^2_{sol} \sim 10^{-4}$ eV$^2$, this approximation is
not reliable for certain oscillation channels such as $\nu_\mu \to \nu_e$.
Here we are referring to CP-conserving quantities; the one-$\Delta m^2$
approximation is, of course, not used for calculating CP-violating quantities
since neglecting $\Delta m^2_{sol}$ is equivalent to setting two of the
neutrino masses equal (in the standard three-active-flavor context), which
allows one to rotate away the CP-violating phase that would appear in neutrino
oscillation experiments and hence renders CP-violating quantities trivially
zero.  Since the values of $\sin^2(2\theta_{13})$ and $\Delta m^2_{sol}$ for
which the one-$\Delta m^2$ approximation breaks down are in the range of
interest for future experimental searches for $\nu_\mu \to \nu_e$ via both
conventional neutrino beams generated by pion decay and via neutrino beams from
neutrino ``factories'' based on muon storage rings, this complicates the
analysis of the sensitivity and data analysis from these experiments.  Many
fits have been performed of solar and atmospheric data, e.g.
\cite{fitsol,othersol}, \cite{bgg}-\cite{bds}.  One salient result is that in
Ref. \cite{foglidem}, the usual fit to the SuperK atmospheric neutrino data
with a single $\Delta m^2_{atm} \simeq 3 \times 10^{-3}$ eV$^2$ and $\sin^2
(2\theta_{atm}) = 1$ is compared with a very different fit with two equal mass
squared differences, $\Delta m^2_{32}=\Delta m^2_{21}=0.7 \times 10^{-3}$
eV$^2$, and it is argued that although the $\chi^2$ for the latter is worse, it
is still an acceptable fit.  This suggests that one should carefully assess
corrections to the one-$\Delta m^2$ approximation in studies of neutrino
oscillations.

\section{Generalities on Neutrino Mixing and Oscillations} 

\subsection{Mixing Matrix and Oscillation Probabilities} 

In this section we briefly record some standard formulas for neutrino
oscillations that we shall use.  In the framework of three active neutrinos,
the unitary transformation relating the mass eigenstates $\nu_i$, $i=1,2,3$, to
the weak eigenstates $\nu_a$ is given by $\nu_a = \sum_{i=1}^3 U_{ai} \nu_i$
where the lepton mixing matrix is
\beqs
U &=& R_{23}KR_{13}K^*R_{12}K'=\\
&=& 
\pmatrix{c_{12} c_{13} & s_{12}c_{13} & s_{13} e^{-i\delta} \cr 
-s_{12}c_{23}-c_{12}s_{23}s_{13}e^{i\delta}
& c_{12}c_{23}-s_{12}s_{23}s_{13}e^{i\delta} & s_{23}c_{13} \cr 
s_{12}s_{23}-c_{12}c_{23}s_{13}e^{i\delta}
&-c_{12}s_{23}-s_{12}c_{23}s_{13}e^{i\delta} & c_{23}c_{13}}\!K'
\label{u}
\eeqs
Here $R_{ij}$ is the rotation matrix in the $ij$ subspace,
$c_{ij}=\cos\theta_{ij}$, $s_{ij}=\sin\theta_{ij}$, $K=diag(e^{-i\delta},1,1)$
and $K'=diag(1,e^{i\delta_1},e^{i\delta_2})$ involves further possible phases
(due to Majorana mass terms) that do not contribute to neutrino oscillations
(as can be seen from the invariance of the quantity $K_{ab,ij}$ below under
neutrino field rephasings).  One can take $\theta_{ij} \in [0,\pi/2]$ with
$\delta \in [0,2\pi)$.  The rephasing-invariant measure of CP violation
relevant to neutrino oscillations is given by the Jarlskog invariant \cite{j}
determined via the product $Im(U_{ij}U_{kn}U^*_{in}U^*_{kj})$, 
\beq
J = \frac{1}{8}\sin(2\theta_{12})
\sin(2\theta_{23})\sin(2\theta_{13})\cos \theta_{13} \sin\delta 
\label{j}
\eeq

In vacuum, the probability that a weak neutrino eigenstate
$\nu_a$ becomes $\nu_b$ after propagating a distance $L$ (assuming that 
$E >> m(\nu_i)$ and the propagation of the mass eigenstates is coherent) is 
\beqs
P(\nu_a \to \nu_b) &=& \delta_{ab} - 4 \sum_{i>j=1}^3 
Re(K_{ab,ij}) \sin^2 \phi_{ij} 
\nonumber\\&+& 4 \sum_{i>j=1}^3 Im(K_{ab,ij})
 \sin \phi_{ij} \cos \phi_{ij} 
\label{pab}
\eeqs
where
\beq
K_{ab,ij} = U_{ai}U^*_{bi}U^*_{aj} U_{bj}
\label{k}
\eeq
\beq
\Delta m_{ij}^2 = m(\nu_i)^2-m(\nu_j)^2
\label{delta}
\eeq
and
\beq
\phi_{ij} = \frac{\Delta m^2_{ij}L}{4E}
\label{phiij}
\eeq 
We recall that for the CP-transformed reaction $\bar\nu_a \to \bar\nu_b$ and
the time-reversed reaction $\nu_b \to \nu_a$, the oscillation probabilities are
given by eq. (\ref{pab}) with the sign of the $Im(K_{ab,ij})$ term reversed.
Further, by CPT, $P(\bar\nu_b \to \bar\nu_a) = P(\nu_a \to \nu_b)$ so that, in
particular, $P(\bar\nu_a \to \bar\nu_a) = P(\nu_a \to \nu_a)$.  It is
straightforward to substitute the elements of the lepton mixing matrix
(\ref{u}) and evaluate the general formula (\ref{pab}) for each of the relevant
transitions.

For the special case $P(\nu_a \to \nu_a)$, eq. (\ref{pab}) simplifies to
\beq
P(\nu_a \to \nu_a) = 1-4\sum_{i > j =1}^3 |U_{ai}|^2|U_{aj}|^2
\sin^2 \phi_{ij}
\label{paa}
\eeq
We recall the elementary identity
\be
\Delta m^2_{32} + \Delta m^2_{21} + \Delta m^2_{13} = 0
\ee
so that in general, three-flavor vacuum oscillations depend on the four angles
$\theta_{12}$, $\theta_{23}$, $\theta_{13}$, $\delta$ and two $\Delta m^2$'s,
which can be taken to be $\Delta m^2_{32}$ and $\Delta m^2_{21}$.  The
currently favored regions for $\theta_{21}$ and $\Delta m^2_{21}$ are
determined primarily by the solar neutrino
data.  Here one can take $\Delta m^2_{21} > 0$ with $\theta_{12} \in [0,\pi/2]$
\cite{darkside}.  To distinguish between the first and second octants, the
parameter regions allowed by these fits to the solar data can be expressed in
terms of $\Delta m^2_{21} > 0$ and $\tan^2\theta_{21}$.

The commonly used one-$\Delta m^2$ approximation is then based on the hierarchy
\beq
\Delta m^2_{sol} \equiv \Delta m^2_{21} 
\ll |\Delta m^2_{31}| \approx |\Delta m^2_{32}|
\equiv |\Delta m^2_{atm}|
\label{hierarchy}
\eeq
However, as mentioned above, the solar data itself or in combination with
atmospheric and reactor data allows for rather large values of $\Delta
m^2_{sol}$, up to $\sim 2 \times 10^{-4}$ eV$^2$ or even somewhat higher.  

\subsection{Two-flavor Oscillations} 

In the case of oscillations of two flavors, the oscillation probability in 
vacuum is, in an obvious notation, 
\beq
P(\nu_a \to \nu_b) = \sin^2(2\theta)\sin^2 \phi_{ij}
\label{twofl}
\eeq

\subsection{Three-Flavor Oscillations with One-$\Delta m^2$ Dominance}

With the hierarchy (\ref{hierarchy}), one has the following approximate
formulas for vacuum oscillation probabilities relevant for experiments with 
reactor antineutrinos, atmospheric neutrinos and CP-conserving effects in 
terrestial long-baseline oscillation studies: 
\beqs
P(\nu_\mu \to \nu_\tau) & = & 4|U_{33}|^2|U_{23}|^2 \sin^2 \phi_{atm} \cr\cr
& = & \sin^2(2\theta_{23})\cos^4 \theta_{13} \sin^2 \phi_{atm}
\label{pnumunutau}
\eeqs
\beqs
P(\nu_\mu \to \nu_e) & = & 4|U_{13}|^2 |U_{23}|^2 \sin^2 \phi_{atm} \cr\cr
     & = & \sin^2(2\theta_{13})\sin^2 \theta_{23} \sin^2 \phi_{atm} 
\label{pnumunue}
\eeqs
\beqs
P(\nu_e \to \nu_\tau) & = & 4|U_{33}|^2 |U_{13}|^2 \sin^2 \phi_{atm} \cr\cr
      & = & \sin^2(2\theta_{13})\cos^2 \theta_{23} \sin^2 \phi_{atm} 
\label{pnuenutau}
\eeqs
Since this one-$\Delta m^2$ approximation removes CP-violating terms, it also
implies that the oscillation probabilities for the CP-transformed and
T-reversed transitions are equal to the probability for the original
transition, $P(\bar\nu_a \to \bar\nu_b)_{1DA} = P(\nu_b \to \nu_a)_{1DA} = 
P(\nu_a \to \nu_b)_{1DA}$. 

For the analysis of data on reactor antineutrinos, i.e. tests of 
$\bar\nu_e \to \bar\nu_e$, the two-flavor mixing expression is 
$P(\bar\nu_e \to \bar\nu_e)=1-P(\bar\nu_e \not \to \bar\nu_e)$, where 
\beq
P(\bar\nu_e \not \to \bar\nu_e) = \sin^2(2\theta_{reactor})
\sin^2 \Bigl (\frac{\Delta m^2_{reactor}L}{4E} \Bigr )
\label{preactor}
\eeq
where $\Delta m^2_{reactor}$ is the squared mass difference relevant for 
$\bar\nu_e \to \bar\nu_x$.  Combining (\ref{pnumunue}) and (\ref{pnuenutau}),
we have, in this approximation, 
\beq
\theta_{reactor} = \theta_{13} \ , \quad \Delta m^2_{reactor} = 
\Delta m^2_{32}
\label{thetareactor}
\eeq

For the analysis of atmospheric data with the transition favored by the current
data, letting 
\beq
P(\nu_\mu \to \nu_\tau) = \sin^2(2\theta_{atm})
\sin^2 \Bigl (\frac{\Delta m^2_{atm}L}{4E} \Bigr )
\label{patm}
\eeq
one has, using (\ref{pnumunutau}), 
\beq
\sin^2(2\theta_{atm}) \equiv \sin^2(2\theta_{23}) \cos^4 \theta_{13}
\label{sinatm}
\eeq
and $\Delta m^2_{atm}$ as given in (\ref{hierarchy}).  Since the best fit value
in the SuperK experiment is $\sin^2(2\theta_{atm})=1$, it follows that
\beq
\theta_{13} << 1
\label{theta13limit}
\eeq
and hence
\beq
\theta_{atm} \simeq \theta_{23}
\eeq

For the K2K experiment, using $\theta_{13} << 1$, one has
\beq
P(\nu_\mu \to \nu_\mu) \simeq 1 - \sin^2(2\theta_{23}) \sin^2 \phi_{atm}
\label{pk2k}
\eeq

All of these vacuum oscillation probabilities are independent of the sign of
$\Delta m^2_{atm}$, just as in the two-flavor vacuum case. However, the
symmetry $\theta \to \pi/2- \theta$ of the two-flavor vacuum case is no longer
present.  For $\theta_{13}$, one can immediately infer that this angle is near
0 rather than near to $\pi/2$ from the fit to the atmospheric data, as noted
above.  For $\theta_{23}$, the transformation $\theta_{23} \to \pi/2 -
\theta_{23}$ leaves the expression for $P(\nu_\mu \to \nu_\tau)$ in
(\ref{pnumunutau}) invariant and interchanges the values of the oscillation
probabilities $P(\nu_e\to\nu_\mu)$ and $P(\nu_e\to\nu_\tau)$.  Because we know
that the value of $\theta_{23}$ is close to $\pi/4$, this interchange does not
make a large change in these probabilities $P(\nu_e\to\nu_\mu)$ and
$P(\nu_e\to\nu_\tau)$.  The atmospheric data places an upper bound on the
transition $P(\nu_\mu \to \nu_e)$, and the fact that this is small is implied
by the fact that $\theta_{13} << 1$, so that this upper bound does not
determine how large the $\sin^2\theta_{23}$ factor in (\ref{pnumunue}) is and
hence whether $\theta_{23}$ is slightly below or slightly above $\pi/4$.

\section{Generalized Analysis of Reactor Antineutrino Data}

The general three-flavor, two-$\Delta m^2$ (i.e., two independent $\Delta m^2$)
formula for antineutrino survival probability that is measured in reactor
experiments such as G\"osgen, Bugey, Chooz, Palo Verde, and KamLAND is
$P(\bar\nu_e \to \bar\nu_e) = 1-P(\bar\nu_e \not\to \bar\nu_e)$, where, using
(\ref{paa}), we have (note that by CPT, $P(\bar\nu_e \not \to \bar\nu_e)=
P(\nu_e \not \to \nu_e)$)
\beqs
P(\bar\nu_e \not \to \bar\nu_e) & = & 4\sum_{i>j=1}^3 
|U_{ei}|^2 |U_{ej}|^2 \sin^2 \phi_{ij} \cr\cr
& = & \sin^2(2\theta_{13}) \sin^2 \theta_{12} \sin^2 \phi_{32} + 
      \sin^2(2\theta_{13}) \cos^2 \theta_{12} \sin^2 \phi_{31} \cr\cr & + & 
      \sin^2(2\theta_{12}) \cos^4 \theta_{13} \sin^2 \phi_{21}
\label{pnuenue}
\eeqs
Matter effects are negligible for these experiments.  Let us consider
the results from the Chooz experiment, since these place the most stringent
constraints on the relevant parameters.  This experiment obtained the result
\cite{chooz} 
\beq
R = \frac{N_{measured}}{N_{calculated}} = 1.01 \pm 0.028 (stat.) \pm 0.027 
(sys.) 
\eeq
 From this and the agreement between the measured and expected positron 
energy spectra, this experment set the limit
\beq
P(\bar\nu_e \not \to \bar\nu_e) < \epsilon_{Ch}
\label{eps}
\eeq
where $\epsilon_{Ch} \simeq 0.05$.  Within the usual context of two-flavor 
mixing described by (\ref{preactor}), the Chooz experiment reported the 
90 \% confidence level (CL) limits 
\beq
\sin^2 2\theta_{reactor} < 0.1  \quad {\rm for \ large} \quad \Delta
m^2_{reactor} 
\label{sin2thetachooz}
\eeq
and
\beq
|\Delta m^2_{reactor}| < 0.7 \times 10^{-3} \ \ {\rm eV}^2 \quad {\rm for} \ \
\sin^2 2\theta_{reactor}=1
\label{deltamsqchooz}
\eeq
In the one-$\Delta m^2$ approximation, the first two terms of
eq. (\ref{pnuenue}) combine to make the term $\sin^2 2 \theta_{13}\sin^2
\phi_{atm}$ so that 
\beq
\sin^2 2\theta_{13} = \sin^2 2\theta_{reactor} < 0.1
\label{sin2theta13chooz}
\eeq

Since each of the three terms $T_i$ in the general equation (\ref{pnuenue}) is
positive-definite, we have, in an obvious notation, $T_i < \epsilon_{Ch}$.
Applying this to the third term and using the fact that $\theta_{13} << 1$, we
obtain an upper bound on $\sin^2 2\theta_{12} \sin^2 \phi_{21}$.  From the plot
of the Chooz excluded region \cite{chooz}, we infer the pair of bounds
\beq
\Delta m^2_{21} < \cases{ 0.7 \times 10^{-3} \ {\rm eV}^2 & for 
                                              $\sin^2 2\theta_{12}= 1$ \cr
                    &   \cr
                          0.9 \times 10^{-3} \ {\rm eV}^2  & for 
                                              $\sin^2 2\theta_{12}= 0.8$} 
\label{propfac}
\eeq
where the second upper bound applies for the central value of
$\sin^2 2\theta_{12}=0.8$ in the LMA solution, 
\beq
{\rm LMA(central)}: \quad \tan^2 \theta_{12} = 0.4 \ , \qquad
\Delta m^2_{21} = 0.5 \times 10^{-4} \ \ {\rm eV}^2
\label{lma}
\eeq
For the central values of the LMA solution in (\ref{lma}), using $L = 1$ km and
a typical $\bar\nu_e$ energy $E \sim 4$ MeV, the third term in (\ref{pnuenue})
has a value of about $1.3 \times 10^{-5}$, which is negligibly small.  This
increases to about $2 \times 10^{-4}$ for $\Delta m^2_{21} = 2 \times 10^{-4}$
eV$^2$, which is again negligible compared with the range of $\bar\nu_e$
disappearance, $\sim 0.05$, probed by Chooz.

Next, we consider the KamLAND long-baseline reactor experiment, which will use
a liquid scintillator detector in the Kamioka mine to measure $\bar\nu_e p \to
e^+ n$ events initiated by $\bar\nu_e$'s from a number of power reactors and
thereby test the LMA solution to the solar neutrino deficit and is expected to
begin data-taking in 2001 \cite{kamland}. The power reactors are located at
various distances from 140 km to 200 km from Kamioka.  It has been estimated
that, in the absence of oscillations, a total of 1075 $\bar\nu_e p \to e^+ n$
events per kton-yr will be recorded, and of these, 348 events per kton-yr will
arise from the single most powerful reactor, the Kashiwazaki 24.6 GW (thermal)
facility a distance $L=160$ km away \cite{kamland}.  For the conditions of this
experiment, $|\phi_{3j}| >> 1$ for $j=1,2$, so that the $\sin^2 \phi_{3j}$
factors average to 1/2 over the $\bar\nu_e$ energy spectra from the reactors,
and hence (\ref{pnuenue}) reduces to 
\beq
P(\bar\nu_e \not \to \bar\nu_e)_{Kam.} = \frac{1}{2}\sin^2(2\theta_{13}) + 
\sin^2(2\theta_{12}) \cos^4 \theta_{13} \sin^2 \phi_{21}
\label{pnuenuekam}
\eeq
The first term has a maximum value of 0.05.  For the central LMA values in
(\ref{lma}), the second term has a value of approximately 0.1 (almost
independently of $\theta_{13}$, given the bound (\ref{sin2theta13chooz})).
Thus, if $\sin^2(2\theta_{12})$ and $\Delta m^2_{21}$ are characterized by the
LMA solution and if $\sin^2(2\theta_{13})$ is near to its current upper bound,
then the two terms in eq. (\ref{pnuenuekam}) would make contributions that
differ only by about a factor of 2.  One can thus distinguish several possible
outcomes for the KamLAND experiment:
\begin{itemize}

\item 

 If this experiment sees a signal for oscillations of reactor $\bar\nu_e$'s
with $P(\bar\nu_e \not \to \bar\nu_e) > 0.05$, this implies that there is at
least some contribution to this signal from the second term.

\item 

In general, if $P(\bar\nu_e \not \to \bar\nu_e)$ is nonzero, then, from the
overall deficiency in the rate, one would not be able to determine the relative
contributions of each term in (\ref{pnuenuekam}).  Instead, for one pathlength,
$L$, one would have to perform a three-parameter fit involving the parameters
$c_j$, $j=1,2,3$, where $c_1$ is a constant, representing the first term in
(\ref{pnuenuekam}), and $c_j$, $j=2,3$ enter as $c_2 \sin^2(c_3 E/(4L))$,
representing the second term in (\ref{pnuenuekam}).  Since the KamLAND detector
is sensitive to $\bar\nu_e$'s from a number of different reactors at different
distances, the actual fit to the data would be more complicated than this, but
the oscillations would not, in general, be adequately described by a
two-flavor, one-$\Delta m^2$ formula, and eq. (\ref{pnuenuekam}) would apply
for the three-flavor, two-$\Delta m^2$ analysis, given the size of $\Delta
m^2_{atm}$ inferred from the atmospheric data.  Clearly, the actual results
extracted from the data will depend on simulations that involve the
determination and subtraction of reactor backgrounds \cite{bk}.

\item 

Finally, if the KamLAND experiment sees no signal for oscillations and sets the
limit $P(\bar\nu_e \to \bar\nu_e) < \epsilon_{KL}$, then since each of the
terms in (\ref{pnuenuekam}) is positive-definite, one will have the bounds 
$\sin^2(2\theta_{13}) < 2\epsilon_{KL}$ and the usual excluded-region plot for
the contribution of the second term.  This will depend on the statistical
uncertainties and the backgrounds and resultant systematic uncertainties, but,
roughly speaking, it will have asymptotes $\sin^2(2\theta_{12}) <
2\epsilon_{KL}$ if $\phi_{21}$ is assumed to be large enough so that 
$\sin^2 \phi_{21}$ averages to 1/2 over the reactor $\bar\nu_e$ energy spectra,
and a corresponding bound of $\sin^2 \phi_{21} < \epsilon_{KL}$ for maximal
mixing, $\sin^2 2\theta_{12}=1$ (given that one knows that the $\cos^4
\theta_{13}$ factor is very close to unity).  

\end{itemize}

\section{Generalized Analysis of a Long-Baseline Experiment to Measure 
$\nu_\mu \to \nu_{\lowercase{e}}$}

In this section we shall discuss the general three-flavor, two-$\Delta m^2$
analysis of long-baseline accelerator experiments to measure $\nu_\mu \to
\nu_e$ using conventional beams and $\nu_e \to \nu_\mu$ or its conjugate using
beams from a possible future neutrino factory based on a muon storage ring.
There are several long-baseline accelerator experiments under construction to
continue the study of neutrino oscillations after the pioneering work of the
K2K experiment.  These include the MINOS experiment from Fermilab to the Soudan
mine, with $L = 730$ km, using a far detector of steel and scintillator and a
neutrino flux peaked at $E \sim 3$ GeV \cite{minos}.  In Europe, a program is
underway to use a neutrino beam with $E \sim 20$ GeV from CERN a distance
$L=730$ km to the Gran Sasso deep underground laboratory, involving the OPERA
experiment and also plans for a liquid argon detector \cite{cngs}.  Third, the
JHF-SuperK neutrino oscillation experiment will use a $\nu_\mu$ beam from the
0.75 MW Japan Hadron Facility (JHF) High Intensity Proton Accelerator (HIPA) in
Tokai, travelling a distance $L=295$ km to Kamioka \cite{jhf}.  In a first
stage, this would use SuperK as the far detector; a possibility that is
discussed for a second stage involves an upgrade of JHF to 4 MW and the
construction of a 1 Mton water Cherenkov far detector (denoted
HyperKamiokande).  The JHF-SuperK collaboration has stated that one of the
goals of its first phase is to search for $\nu_\mu \to \nu_e$ oscillations down
to the level $P(\nu_\mu \to \nu_e) \sim 0.003$ by taking advantage of the
excellent particle identification ability and energy resolution of SuperK for
electrons and muons \cite{jhf}.  This is the sensitivity for a narrow-band beam
with $E=0.7$ GeV, which, assuming that $|\Delta m^2_{32}|=3 \times 10^{-3}$
eV$^2$, maximizes the factor $\sin^2 \phi_{32}$; the estimated sensitivity for
a wide-band beam with $E$ peaked at about 1.1 GeV is $P(\nu_\mu \to \nu_e) \sim
10^{-2}$ \cite{jhf}.  This type of search will be pursued to some level also by
the other long-baseline experiments.  A number of other possible long-baseline
$\nu_\mu \to \nu_e$ oscillation experiments using intense conventional 
neutrino beams have been considered, with a variety of pathlengths 
\cite{superbeam}. 

A different approach that has been considered in detail is that of a neutrino
``factory'', in which one would obtain a very intense beam of $\nu_\mu$ and
$\bar\nu_e$'s from the decays of $\mu^-$'s in a muon storage ring in the form
of a racetrack or bowtie, and similarly a beam of $\bar\nu_\mu$ and $\nu_e$'s
from stored $\mu^+$'s.  These beams would have a definite and precisely
understood flavor composition and would make possible neutrino oscillation
searches using very long-baselines of order 3000 km \cite{anl}-
\cite{nufactmeasurements}.  Typical design parameters are $E=20$ GeV for the
stored $\mu^\pm$ energy and $L = 10^{20}$ $\mu$ decays per Snowmass year
($10^7$ sec).  With a stored $\mu^-$ beam, say, one would carry out a
measurement of the $\nu_\mu \to \nu_\mu$ survival probability via the charged
current reaction yielding a final state $\mu^-$ and an appearance experiment
with $\bar\nu_e \to \bar\nu_\mu$ yielding a final state $\mu^+$, a so-called
wrong-sign muon signature.  It has been estimated that with a moderate-level
neutrino factory, one could search for $\nu_e \to \nu_\mu$ or its conjugate
reaction down to the level $\sin^2(2\theta_{13}) \sim 3 \times 10^{-4}$
\cite{fnal}.  For such long pathlengths, matter effects are important
\cite{nufactmeasurements} and can be used to get information on the sign of
$\Delta m^2_{32}$.  It may also be possible to measure leptonic CP violation
using either a conventional beam or a beam from a neutrino factory.

At the levels of $\sin^2 2\theta_{13}$ that will be probed, the one-$\Delta
m^2$ approximation used in many planning studies may well be inadequate, and
one should use a more general theoretical framework.  The full expression for
the $\nu_\mu \to \nu_e$ oscillation probability in vacuum (matter effects are
discussed below) is obtained in a straighforward manner from the formulas
(\ref{pab}) and (\ref{u}) and has the form, in a compact notation,
\beqs
P(\nu_\mu\to \nu_e)&=&2 \sin(2\theta_{13}) s_{23}c_{13}s_{12}
(s_{12}s_{23}s_{13}-c_{12}c_{23} c_\delta)\sin^2\phi_{32}
+\nonumber \\
&+&2 \sin(2\theta_{13}) s_{23}c_{13}c_{12}
(c_{12}s_{23}s_{13} +s_{12}c_{23} c_\delta)\sin^2\phi_{31}
-\nonumber \\
&-& 2 \sin(2\theta_{12}) c_{13}^2 \biggl [ 
s_{12} c_{12}(s_{13}^2 s_{23}^2-c_{23}^2)+s_{13} s_{23} c_{23} 
(s_{12}^2-c_{12}^2)c_\delta \biggr ] \sin^2\phi_{21}
\nonumber \\
&+& \frac{1}{2} \sin(2\theta_{12}) \sin(2\theta_{13})\sin(2\theta_{23})c_{13}
s_\delta
\bigg[ \sin \phi_{32}
\cos \phi_{32}
-\nonumber \\
&-&\sin\phi_{31}
\cos\phi_{31}
+\sin\phi_{21}
\cos\phi_{21}
\bigg]
\label{pnumunuefull}
\eeqs
(As in (\ref{pab}), $P(\bar\nu_\mu \to \bar\nu_e)$ and $P(\nu_e \to \nu_\mu)$
are given by (\ref{pnumunuefull}) with the sign of the $\sin\delta$ term
reversed.)  Now for sufficiently small $\Delta m^2_{21}$, as would be true in
the solar neutrino fits with SMA, LOW, or $\Delta m^2_{21} \sim 10^{-9}$ eV$^2$
vacuum oscillations, and sufficiently large $\sin^2(2\theta_{13})$, subject to
the constraint (\ref{sin2theta13chooz}), the full eq. (\ref{pnumunuefull})
reduces to (\ref{pnumunue}), in which the oscillation is driven by the terms
involving $\sin^2 \phi_{atm} = \sin^2 \phi_{32} \simeq \sin^2 \phi_{31}$.
However, if $\Delta m^2_{21}$ and $\sin^2 2\theta_{21}$ are at the upper end of
the LMA region, then the one-$\Delta m^2$ approximation can break down.  As a
numerical example, one can consider the parameter set $\sin^2 2
\theta_{12}=0.8$, $\Delta m^2_{21} = 2 \times 10^{-4}$ eV$^2$, $\sin^2
2\theta_{13} = 0.01$, $\delta = \pi/6$, with the usual central SuperK values
$\Delta m^2_{32}=3 \times 10^{-3}$ eV$^2$ and $\sin^2 2\theta_{23}=1$.
Further, take the JHF-SuperK pathlength $L=295$ km and narrow-band-beam energy
$E=0.7$ GeV, and label this total set of parameters as set (a).  Then, if one
were to evaluate the $\nu_\mu \to \nu_e$ oscillation probability using the
one-$\Delta m^2$ approximation, again denoted 1DA, eq. (\ref{pnumunue}), one 
would obtain
\beq
P(\nu_\mu \to \nu_e) = 5.0 \times 10^{-3} \quad {\rm for \ set \ (a) \ with 
\ 1DA} 
\eeq
However, correctly including the contribution from the term involving $\sin^2
\phi_{21}$, using the full expression (\ref{pnumunuefull}), one gets an
oscillation probability that is more than twice as large as the one predicted
by the one-$\Delta m^2$ approximation: 
\beq
P(\nu_\mu \to \nu_e) = 1.4 \times \times 10^{-2} \quad {\rm for \ set \ (a)}
\eeq
This clearly shows that for experimentally allowed input parameters involving
the LMA solar fit, and in particular, for a value of $\sin^2 2\theta_{13}$ that
can be probed by the JHF-SuperK experiment and others that could achieve
comparable sensitivity, the one-$\Delta m^2$ approximation may not be valid.
Thus, it is important that the KamLAND experiment will test the LMA and
anticipates that, after about three years of running, it will be sensitive to
the level $\Delta m^2_{sol} \lsim 10^{-5}$ eV$^2$ \cite{kamland}.  This
information should therefore be available by the commissioning of JHF in 2007.
The adequacy of the three-flavor theoretical framework will also be tested by
the miniBOONE experiment within this period.  If, indeed, the LMA parameter set
is confirmed by KamLAND, then it may well be necessary to take into account
three-flavor oscillations involving two independent $\Delta m^2$ values in the
data analysis for the JHF-SuperK experiment and other $\nu_\mu \to \nu_e$
neutrino oscillation experiments that will achieve similar sensitivity.  This
point is thus certainly also true for long-baseline experiments with a neutrino
factory measuring $\nu_e \to \nu_\mu$, $\bar\nu_e \to \bar\nu_\mu$
oscillations, since they anticipate sensitivity to values of $\sin^2 2
\theta_{13}$ that are substantially smaller than the level to which the
JHF-SuperK collaboration will be sensitive, and as one decreases $\theta_{13}$
with other parameters held fixed, the $\sin^2 \phi_{21}$ corrections to the
one-$\Delta m^2$ approximation become relatively more important.  

In passing, we observe that in the limit $\theta_{13} \to 0$,
eq. (\ref{pnumunuefull}) reduces to 
\beq
P(\nu_\mu \to \nu_e) = \sin^2(2\theta_{12})\cos^2 \theta_{23} \sin^2 \phi_{21}
\quad {\rm for} \quad \theta_{13}=0
\label{pnumunuet130}
\eeq 
In this limit, the term involving $\sin^2 \phi_{21}$, rather than the terms
involving $\sin^2 \phi_{32}$ or $\sin^2 \phi_{31}$, are driving the $\nu_\mu
\to \nu_e$ oscillations.

\section{$\nu_\mu \to \nu_\mu$ Disappearance Experiments}

All long-baseline accelerator neutrino experiments, including K2K, MINOS, CNGS,
JHF-SuperK, and other possible ones such as CERN-Frejus and those that might
involve UNO and/or a neutrino factory, will perform a measurement of the
$\nu_\mu \to \nu_\mu$ survival probability.  The one-$\Delta m^2$ approximation
yields the result
\beqs
P(\nu_\mu \not \to \nu_\mu) & = & 4(1-|U_{\mu 3}|^2)|U_{\mu 3}|^2 
\sin^2 \phi_{32} \cr\cr
& = & \biggl [ \sin^2 (2\theta_{23})\cos^2 \theta_{13} + 
\sin^2 (2\theta_{13}) \sin^4 \theta_{23} \biggr ] \sin^2 \phi_{32}
\label{pnumunumu1da}
\eeqs
Since SuperK infers a maximal $\nu_\mu \to \nu_\tau$ oscillation to fit its
atmospheric neutrino data, and since this implies that $\theta_{13} << 1$, the
second term in (\ref{pnumunumu1da}) is quite small compared to the first.  As a
numerical example, for $\sin^2 2\theta_{13}=0.01$ and $\theta_{23}=\pi/4$, the
ratio of the second to the first term in (\ref{pnumunumu1da}) is $2.5 \times
10^{-3}$.  The one-$\Delta m^2$ approximation is a very good one for this
transition; for experiments such as MINOS and JHF-SuperK, the relative 
corrections are typically of order $\lsim O(10^{-2})$.

\section{$\nu_{\lowercase{e}} \to \nu_\tau$}

This transition is more difficult to measure than $\nu_\mu \to \nu_e$ since (a)
the optimal neutrino energy to maximize the oscillation factor is below $\tau$
threshold, and (b) even if this were not the case, the $\tau$ is not observed
directly.  For completeness, however, it should be noted that again the term
retained in the usual one-$\Delta m^2$ approximation, (\ref{pnuenutau}) may not
be larger than the term that would describe this transition if $\theta_{13}=0$,
namely 
\beq
P(\nu_e \to \nu_\tau) = \sin^2(2\theta_{12})\sin^2 \theta_{23}\sin^2 \phi_{21}
\quad {\rm for} \quad \theta_{13}=0
\label{pnuenutaut130}
\eeq
This is the same as the expression for $P(\nu_e \to \nu_\mu)= P(\nu_\mu \to
\nu_e)$, eq. (\ref{pnumunuet130}) under the same assumption, $\theta_{13}=0$
with the interchange of $\cos^2\theta_{23}$ and $\sin^2\theta_{23}$.

\section{Matter Effects for Neutrino Oscillations with Two Relevant 
$\Delta {\lowercase{m}}^2$ Scales} 

In many experiments matter effects can be relevant.  This is the case with
solar neutrinos, atmospheric neutrinos, and future possibilities for
$O(10^3)$ km baseline neutrino oscillation experiments using neutrino
factories \cite{fnal}-\cite{cp}. 
In these cases, oscillation probabilities are modified by the
interaction of the neutrinos in the matter: $\nu_\mu$ and $\nu_\tau$ have the
same forward scattering amplitude, via $Z$ exchange, while $\nu_e$ has a
different forward scattering amplitude off of electrons, involving both $Z$ and
$W$ exchange. This leads to a matter-induced oscillation effect when electron
neutrinos are involved in the oscillations.

In this case one needs to solve the evolution equation which includes the
effects of the interactions with matter, which reads (for a generic
two-generation case)
\beq 
i\frac{d}{dx}\pmatrix{\nu_e \cr \nu_\alpha} =
\left(\frac{1}{2E}UM^2U^{\dagger}+V\right)\pmatrix{\nu_e \cr \nu_\alpha} 
\eeq 
where
\beq 
M^2=\pmatrix{m_1^2&0\cr0&m_2^2} , V=\pmatrix{V_e&0\cr0&0},
U=\pmatrix{\cos\theta&\sin\theta\cr -\sin\theta&\cos\theta} 
\eeq 
Here $V=V_e=\sqrt{2}G_FN_e$ where $N_e$ is the electron number density and we
have $\sqrt{2}G_FN_e$ [eV]$=7.6\times 10^{-14} Y_e \rho$ [g/cm$^3$], where
$\rho$ is the mass density and $Y_e$ is the average electron fraction of the
matter. 

Since only relative phases are important for oscillations, we can 
subtract the quantity 
$(1/4E)(m_1^2+m_2^2)+(1/\sqrt{2})G_FN_e$ from the diagonal, 
and the evolution equation becomes:
\beq
i\frac{d}{dx}\pmatrix{\nu_a \cr \nu_b} = \pmatrix {-A(x) & B 
\cr B&A(x)}
\pmatrix{\nu_a\cr\nu_b}
\eeq
with
\beq
A(x)=\frac{\Delta m^2}{4E}\cos(2\theta)-\frac{G_F}{\sqrt2}N_e(x)
\eeq
\beq
B=\frac{\Delta m^2}{4E}\sin(2\theta)
\eeq

For the case of constant density this leads to an oscillation probability
\beq
P(\nu_a \to \nu_b) = \sin^2(2\theta_m)\sin^2(\omega L)
\label{pab2}
\eeq
where 
\beq
\omega=\sqrt{A^2+B^2}=\frac{\Delta m^2}{4E}
\Biggl [ \sin^2(2\theta)+\bigg (\cos(2\theta)-\frac{2\sqrt2 G_F N_e E}
{\Delta m^2}\bigg )^2 \Biggr ]^{1/2} 
\label{omega}
\eeq
gives the effective squared mass difference, divided by $4E$, in matter, and 
$\theta_m$ is the relevant effective mixing angle in matter, specified by 
\beq
\sin^2(2\theta_m)=\frac{\sin^2(2\theta)}{\sin^2(2\theta)+
\Bigl ( \cos(2\theta) - \frac{2\sqrt{2}G_FN_e E}{\Delta m^2} \Bigr )^2 }
\label{thetam}
\eeq
 Thus, the
resonance condition is
\beq
E = 13 \ {\rm GeV}
\biggl (\frac{\Delta m^2}{3 \times 10^{-3} \ {\rm eV}^2 }\biggr ) 
\biggl ( \frac{3 \ {\rm g/cm}^2}{\rho}\biggr ) \biggr (\frac{1/2}{Y_e}\biggr ) 
\cos 2\theta
\label{eresonance}
\eeq
where we have introduced scaling factors normalized by typical values of the
density and the fraction $Y_e=Z/A$ in the upper mantle. 

Letting the vacuum oscillation length $L_{vac}$ be defined as $L_{vac} = 4\pi
E/|\Delta m^2|$, the effective oscillation length $L_m$, in matter, defined by 
$\omega L_m = \pi$, is
\beq
L_m = L_{vac}\Biggl [ \sin^2(2\theta)+\bigg
(\cos(2\theta)-\frac{2\sqrt2 G_F N_e E}
{\Delta m^2}\bigg )^2 \Biggr ]^{-1/2}
\eeq
We recall that, as is evident from these formulas, 
the oscillation probability in matter depends on
the sign of $\cos 2\theta$, i.e., whether $\theta$ is in the first or second
octant, given that one takes $\Delta m^2 > 0$. 

We next recall the formulas for matter effects on oscillation probabilities in
the three-flavor case with the one-$\Delta m^2$ dominance approximation.  Here,
the evolution of the weak eigenstates is given by
\beq
i\frac{d}{dx}\nu = \left(\frac{1}{2E}UM^2U^{\dagger}+V\right)\nu
\label{evmat}
\eeq
where
\beq
\nu=\pmatrix{\nu_e\cr \nu_\mu\cr\nu_\tau}=U \pmatrix{\nu_1\cr \nu_2\cr\nu_3}
\eeq
\beq
M^2=\pmatrix{m_1^2&0&0\cr0&m_2^2&0\cr0&0&m_3^2}
\label{msqared}
\eeq
\beq
V=\pmatrix{\sqrt{2}G_FN_e&0&0\cr0&0&0&\cr0&0&0}
\label{V}
\eeq
Subtracting $m_1^2$ from the diagonal, $M^2$ becomes 
\beq
M^2=\pmatrix{0&0&0\cr0&\Delta m_{sol}^2&0\cr0&0&\Delta m_{atm}^2+
\Delta m_{sol}^2}
\eeq
In order to calculate the oscillation probabilities for long-baseline
terrestrial neutrino oscillation experiments and for analysis of atmospheric 
neutrino data, it is convenient to transform to a new basis 
defined by (e.g. \cite{akh}) 
\beq
\nu=R_{23}\tilde\nu
\eeq
The evolution of $\tilde\nu$ is given by 
\beq
\tilde H = \frac{1}{2E} K R_{13}K^* R_{12}M^2 R_{12}^{\dagger} K 
R_{13}^{\dagger} K^* +V
\label{htilde}
\eeq
In the one-$\Delta m^2$ approximation, this can be reduced to
\beq
\tilde H \simeq \pmatrix{\frac{1}{2E}s_{13}^2\Delta m_{32}^2+\sqrt{2}G_FN_e&0&
\frac{1}{2E}s_{13}c_{13}\Delta m_{32}^2 e^{-i\delta}\cr
0&0&0\cr
\frac{1}{2E}s_{13}c_{13}\Delta m_{32}^2 e^{i\delta}
&0&\frac{1}{2E}c_{13}^2\Delta m_{32}^2}
\label{hs}
\eeq
It can be seen now that in the basis $(\nu_e,\tilde\nu_\mu,\tilde\nu_\tau)$ 
the three-flavor evolution equation decouples, and it is enough to treat the
two-flavor case.  We define $S$ and $P$ by 
\beq
\pmatrix{\nu_e\cr \tilde\nu_\mu\cr\tilde\nu_\tau}(x)=S
\pmatrix{\nu_e\cr\tilde\nu_\mu\cr\tilde\nu_\tau}(0)
\eeq
and
\beq
P\equiv |S_{13}|^2=1-|S_{33}|^2
\eeq
Transforming back to the flavor 
basis $(\nu_e,\nu_\mu,\nu_\tau)$, the probabilities of oscillation become 
\beqs
P(\nu_e\rightarrow\nu_\mu)&=& P(\nu_\mu\rightarrow\nu_e)=s_{23}^2 P\\
P(\nu_e\rightarrow\nu_\tau)&=&c_{23}^2 P\\
P(\nu_\mu\rightarrow\nu_\tau)&=&s_{23}^2c_{23}^2[2-P-2Re(S_{22}S_{33})]
\label{prob}
\eeqs

If in (\ref{hs}) we subtract from the diagonal the quantity 
$D=(1/4E)\Delta m^2_{32}+ (1/\sqrt{2})G_FN_e$, we see that it is then 
necessary to solve the evolution equation for a two-flavor 
neutrino system as in equation (\ref{twofl}), where in $A$ and $B$, 
$\Delta m^2 = \Delta m^2_{32}$ and $\theta =\theta_{13}$. For the case of 
constant density, $S=e^{-i {\tilde H} L}$, so that $S_{33}=
e^{-i DL} (\cos\omega L-i(A/\omega)\sin\omega L)$ 
and $P$ is given by equations (\ref{pab2})-(\ref{thetam}).
Explicitly for $\nu_\mu \to \nu_e$, 
\beq
P(\nu_\mu \to \nu_e) = P(\nu_e \to \nu_\mu) = \sin^2(2\theta_{13,m}) 
\sin^2(\theta_{23})\sin^2( \omega_{32,m}L)
\label{pmuematter}
\eeq

In this case, as in the two-flavor analysis, the interaction with matter makes
the oscillations sensitive to the sign of $\Delta m^2_{atm}$.  For
antineutrinos, the matter potential has the same magnitude and opposite sign,
so one has to solve the same evolution equation where $V$ is replaced by $-V$.
Consequently, if the oscillation probabilities are enhanced by the presence of
the matter for neutrinos, as they are for $\Delta m^2_{atm} > 0$, then they
will be suppressed for antineutrinos and vice versa. From these results it is
evident that $V\rightarrow -V$ is equivalent to $\Delta m^2\rightarrow -\Delta
m^2$. The neutrino factory physics program intends to use this property to
obtain the sign of $\Delta m^2_{32}$ by comparing $P(\nu_e \to \nu_\mu)$ and
$P(\bar\nu_e \to \bar\nu_\mu)$ (e.g., \cite{fnal}).

\begin{figure}                             
\begin{center}
\mbox{\epsfxsize=10truecm
\epsffile{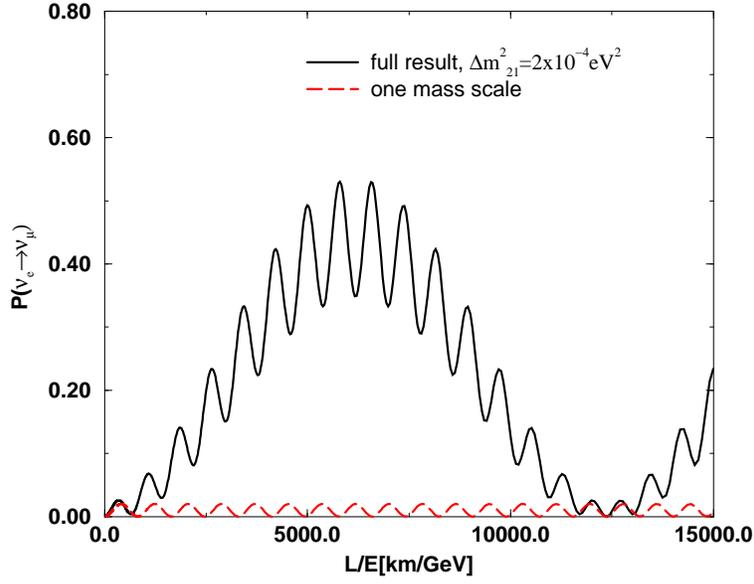}}
\end{center}
\caption{\footnotesize{Plot of $P(\nu_\mu \to \nu_e)$ as a function of $L/E$
for $\sin^2\theta_{23}=1$, $\Delta m^2_{32}=3 \times 10^{-3}$ eV$^2$, 
$\sin^2 2\theta_{12}=0.8$, and $\Delta m^2_{21} = 2\times 10^{-4}$ eV$^2$, 
and $\sin^2 2\theta_{13}=0.04$.}}
\label{fig:muebig}
\end{figure}

Let us now present some numerical results for matter effects in the case of
three-flavor oscillations with two $\Delta m^2$ values.  We concentrate on the
case of large pathlengths and the $\nu_\mu \to \nu_e$ transition relevant to
the existing data on atmospheric oscillations.  For simplicity, we take the
CP-violating phase equal to zero here, but it is straightforward to include it
(see below). These results can also be applied to data on $\nu_e \to \nu_\mu$
that might become available with a possible future neutrino factory.  In
Fig. \ref{fig:muebig} we plot $P(\nu_\mu \to \nu_e)$ as a function of $L/E$ for
$\sin^2 2\theta_{23}=1$, $\Delta m^2_{32}=3 \times 10^{-3}$ eV$^2$, $\sin^2
2\theta_{13}=0.04$, $\sin^2 2\theta_{12}=0.8$, and $\Delta m^2_{21}=2 \times
10^{-4}$ eV$^2$, the upper end of the LMA region. The higher-frequency
oscillations are driven by the terms involving $\sin^2 \phi_{32}$ while the
lower-frequency oscillation is driven by the terms involving $\Delta m^2_{21}$.
The one-$\Delta m^2$ approximation is shown as the dashed curve; of course,
this lacks the low-frequency oscillation component.  One sees that the full
calculation differs strikingly from the result of the one-$\Delta m^2$
approximation.

\begin{figure}                             
\begin{center}
\mbox{\epsfxsize=10truecm
\epsffile{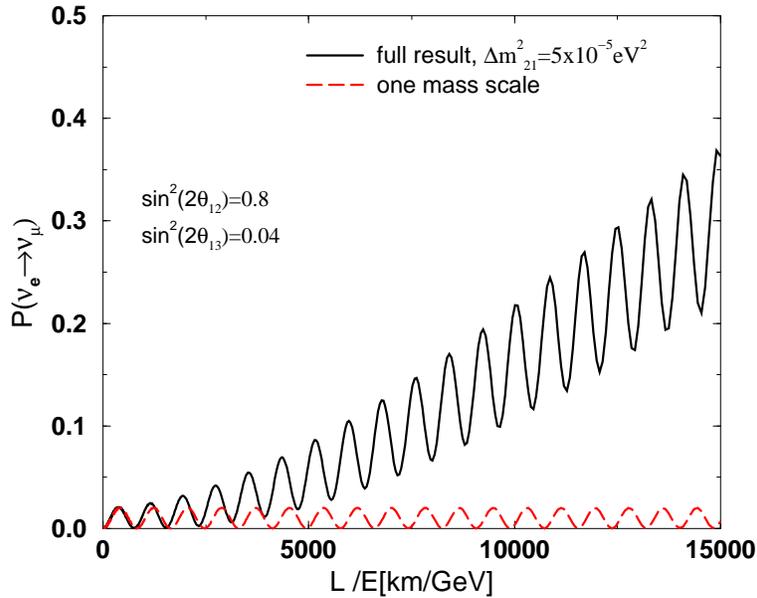}}
\end{center}
\caption{\footnotesize{Plot of $P(\nu_\mu \to \nu_e)$ as a function of $L/E$
for $\sin^2\theta_{23}=1$, $\Delta m^2_{32}=3 \times 10^{-3}$ eV$^2$, central
LMA values $\sin^2 2\theta_{12}=0.8$ and $\Delta m^2_{21}=5 \times 10^{-5}$
eV$^2$, and $\sin^2 2 \theta_{13}=0.04$.}}
\label{fig:muelma}
\end{figure}

Even for the best-fit LMA solution, the effect of $\Delta m^2_{21}$ can be
large for large pathlengths, and this would affect the $\nu_\mu \leftrightarrow
\nu_e$ oscillations in atmospheric neutrino data, as shown in
Fig. \ref{fig:muelma}, for which we take the central values of $\sin^2
2\theta_{21}$ and $\Delta m^2_{21}$ in the LMA fit, (\ref{lma}) and other
parameters the same as in the previous figure.  Note that for the dominant
$\nu_\mu\to\nu_\tau$ transition in the atmospheric neutrinos, $\Delta m^2_{21}$
effects are not so important; this is clear from the fact that this transition
does not directly involve $\nu_e$.

\begin{figure}                             
\begin{center}
\mbox{\epsfxsize=10truecm
\epsffile{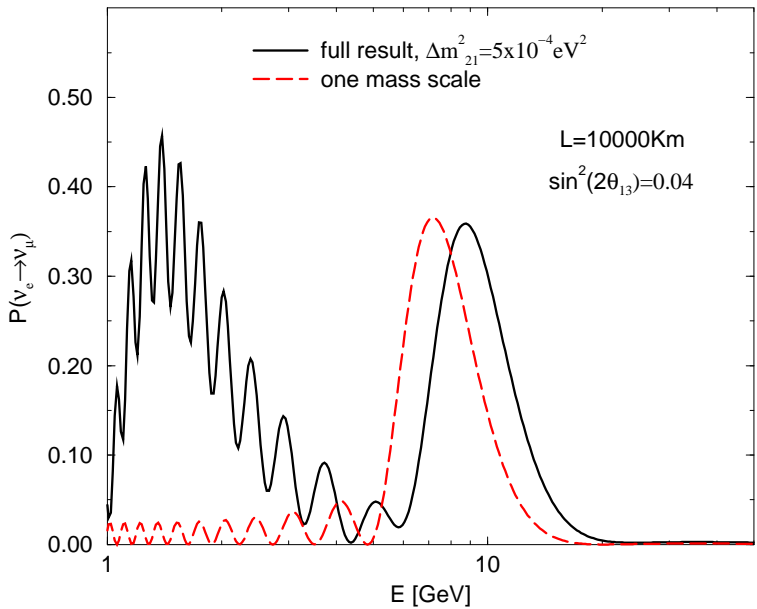}}
\end{center}
\caption{\footnotesize{Plot of $P(\nu_\mu \to \nu_e)$ as a function of $E$
for $\sin^2\theta_{23}=1$, $\Delta m^2_{32}=3 \times 10^{-3}$ eV$^2$, and
$L=10^4$Km, with other input values as shown.  This calculation takes account of
the full density profile of the earth.}}
\label{fig:matter}
\end{figure}

We next show, in Fig. \ref{fig:matter}, the result of integrating (\ref{evmat})
in the full three-flavor mixing scenario and using the actual density profile
of the Earth as given in \cite{prem}. For this figure we use $\sin^2
2\theta_{23}=1$, $\Delta m^2_{32}=3\times 10^{-3}$ eV$^2$,
$\sin^2(2\theta_{12})=0.8$, $\Delta m^2_{21}=5\times 10^{-4}$ eV$^2$, and
$\sin^2(2\theta_{13})=0.04$.  As expected, the $\Delta m^2_{21}$ corrections
are big for low energies and large distances. For the choice of a large
distance, $L=10^4$ km (shown in fig. \ref{fig:matter}), we observe a very
significant difference between the full calculation and the one-$\Delta m^2$
approximation.  This shows again (as does the recent illustrative study in 
Ref. \cite{foglidem}), that it would be valuable to carry out a more
complete analysis of the SuperK and other atmospheric neutrino data with not
just three-flavor oscillations, but also two $\Delta m^2$ values included.
Although the SuperK fit to its data shows that the $\nu_\mu \leftrightarrow
\nu_e$ oscillations make a small contribution, it is important to include this
contribution correctly, and the one-$\Delta m^2$ approximation is not, in
general, reliable for this transition.

\section{CP Violation}

It is well known that there can be substantial leptonic CP violation observable
in neutrino oscillations, in the context of three-flavor mixing and that this
depends on both $\Delta m^2_{32}$ and $\Delta m^2_{21}$ being nonzero.  The
observation of leptonic CP violation is, indeed, a major goal of a neutrino
factory \cite{fnal}-\cite{cp}.  Our main point here pertains to the accuracy
with which input parameters might be known at a time when a neutrino factory
might operate, which, in turn, leads one back to the necessity of a general
three-flavor, two-$\Delta m^2$ analysis of atmospheric neutrino data and data
from studies of $\nu_\mu \to \nu_e$ oscillations with an intense conventional
neutrino beam.  To elaborate on this, we recall that at a neutrino factory, a
potentially promising way to measure CP violation is via the asymmetry
\beq
A_1 = \frac{P(\nu_e \to \nu_\mu) - P(\bar\nu_e \to \bar\nu_\mu)}
{P(\nu_e \to \nu_\mu) + P(\bar\nu_e \to \bar\nu_\mu)}
\label{asym2}
\eeq
This method has two main appeals: (i) one can produce equally intense initial
fluxes of $\nu_e$ and $\bar\nu_e$ by switching the stored beam between $\mu^+$
and $\mu^-$; and (ii) a detector for long-baseline neutrino oscillation
searches with a neutrino factory will have the ability to identify outgoing
$\mu^\pm$'s and measure their electric charges, and its detection efficiency
will be equal for the two signs, so no bias will be introduced in this
measurement.  The complication with this method is that, even in the absence of
any intrinsic CP violation, the asymmetry $A_2$ does not vanish because matter
effects reverse sign between neutrino and antineutrinos, and the earth is not
CP-symmetric.  Therefore, the challenge with this method will be to determine
these matter effects with sufficient accuracy to be able to disentangle them
from the intrinsic CP violation.  An important source of information here will
be the anticipated measurement of $\theta_{13}$ from the JHF-SuperK experiment
on $\nu_\mu \to \nu_e$ oscillations, since matter effects are sensitively
dependent on this parameter \cite{nufactmeasurements}.  As we have shown in a
previous section, for an accurate determination of $\theta_{13}$, the
one-$\Delta m^2$ approximation is not, in general, reliable, especially if
$\sin^2 2\theta_{21}$ and $\Delta m^2_{21}$ are near to their maximal values in
the LMA solution to the solar neutrino deficit. 

It should be noted that there are also plans to try to measure CP violation
with an intense conventional beam, e.g. in the JHF-SuperK experimental program,
by comparing overall rates of $\nu_\mu \to \nu_e$ and $\bar\nu_\mu \to
\bar\nu_e$ and also energy dependences of these signals
\cite{jhf,superbeam} (see also \cite{cp}). For sufficiently large
$\sin^2 2\theta_{13}$, $\sin^2 2 \theta_{12}$, and $\Delta m^2_{21}$, these
methods might provide a way to measure CP violation complementary to that used
with a neutrino factory.  There are a number of challenges with this approach:
(i) the fluxes of $\nu_\mu$ and $\bar\nu_\mu$ are different; (ii) the event
rates would involve several different cross sections, $\nu_e n \to e^- p$ and
$\bar\nu_e p \to e^+ n$ in oxygen nuclei, as well as $\bar\nu_e p \to e^+ n$ on
the hydrogen nuclei in the water molecules; (iii) since it is not possible to
determine the sign of the $e^\pm$ with SuperK, the comparison would have to be
done with data from different periods of operation. Assuming these experimental
challenges can be met, the importance of accurate inputs for the various
parameters is evident from the formula for the CP-violating asymmetry
\beq
A_2 = \frac{P(\nu_\mu \to \nu_e) - P(\bar\nu_\mu \to \bar\nu_e)}
{P(\nu_\mu \to \nu_e) + P(\bar\nu_\mu \to \bar\nu_e)}
\label{asym1}
\eeq
For the JHF-SuperK baseline matter effects are small, so we shall consider the
expression for this asymmetry in vacuum (where $A_2=-A_1$).  Using 
\ba
P(\nu_\mu \to \nu_e) - P(\bar\nu_\mu \to \bar\nu_e) & = & 
4J (\sin 2\phi_{32} + \sin 2\phi_{21} + \sin 2\phi_{13}) \nonumber\\
& = & 16 J \sin\phi_{32}\sin\phi_{31}\sin\phi_{21}
\label{pnuedifvacuum}
\ea
where $J$ was given in eq. (\ref{j}), substituting the expression for 
$P(\nu_\mu \to \nu_e)+P(\bar\nu_\mu \to \bar\nu_e)$ from (\ref{pnumunuefull}), 
and using $\Delta m^2_{21} << \Delta m^2_{32}$, one has 
\beq
A_2 \simeq \frac{\sin(2\theta_{12})\cot \theta_{23} \sin\delta \sin \phi_{21}}{
\sin \theta_{13}}
\eeq
Thus, to extract an accurate measurement of $\sin\delta$, it is clearly
important to have sufficiently accurate inputs for quantities such as
$\theta_{13}$, and this, in turn, motivates the generalization of the
one-$\Delta m^2$ approximation that we have presented above for the analysis of
data on neutrino oscillations.

\section{Conclusions}

In this paper we have performed calculations of neutrino oscillation
probabilities in a three-flavor context, taking into account both $\Delta
m^2_{atm}$ and $\Delta m^2_{sol}$ scales.  We have shown that for values of
$\sin^2(2\theta_{13}) \sim 10^{-2}$ in the range of interest for long-baseline
neutrino oscillation experiments with intense conventional neutrino beams such
as JHF-SuperK and with a possible future neutrino factory, and for $\Delta
m^2_{sol} \sim 10^{-4}$ eV$^2$, the contributions to $\nu_\mu \to\nu_e$
oscillations from both CP-conserving and CP-violating terms involving
$\sin^2(\Delta m^2_{sol}L/(4E))$ can be comparable to the terms involving
$\sin^2(\Delta m^2_{atm}L/(4E))$ retained in the one-$\Delta m^2$
approximation.  Accordingly, we have emphasized the importance of performing a
full three-flavor, two-$\Delta m^2$ analysis of the data on $\nu_\mu \to \nu_e$
oscillations from an experiment with a conventional beam, and on $\nu_e \to
\nu_\mu$, $\bar\nu_e \to \bar\nu_\mu$ oscillations from experiments with a
neutrino factory.  In our study we have included calculations of matter
effects in a three-flavor, two $\Delta m^2$ framework.  Our results also
motivate the analysis of atmospheric neutrino data in this generalized
framework.  

\vspace{6mm}

{\bf Acknowledgments} 

\vspace{3mm} 

This research was supported in part by the U. S. NSF grant PHY-97-22101. 

\vspace{4mm}

\section{Appendix: Analytic Approximation for Three-Flavor Two-$\Delta 
{\lowercase{m}}^2$ Oscillations in Matter}

In matter, for constant density, it is still possible to solve the evolution
equation exactly (\cite{exact}).  However, these results are rather
complicated. For fits to the data or other studies, it can be useful in this
case to use approximate formulas that can still very well describe the
oscillations.

For the LMA solution to the solar data, the effects of $\Delta m^2_{sol}$ can 
no longer be neglected, but there are cases where they are small in 
long-baseline experiments for which the dominant oscillation is controlled by 
$\Delta m^2_{atm}$. In these cases, one can thus treat the effects 
of $\Delta m^2_{sol}$ as a small perturbation. This has been
done in \cite{perturbation}, where the matter effect is also take to be a 
small perturbation. This is a good approximation at short and medium 
distances. Here we give generalized formulas that treat the $\Delta m^2_{sol}$
as a perturbation but allow large matter effects, as is necessary in very long
baseline experiments. 

In order to calculate oscillation probabilities it is now convenient to work 
in a basis defined by:
\be
\nu=U R_{12}^\dagger\nu '
\ee
The evolution of $\nu '$ is given by
\ba
 H'&=&
R_{12}U^\dagger H U R_{12}^\dagger\nonumber\\
&=& \frac{\Delta m_{atm}^2}{2E}\pmatrix{0&0&0\cr0&0&0\cr0&0&1} 
+R_{12}U^\dagger VUR_{12}^\dagger
+\frac{\Delta m^2_{sol}}{2E}R_{12}
\pmatrix{0&0&0\cr0&1&0\cr0&0&1}R_{12}^\dagger\\
&=&\pmatrix{V_e c_{13}^2&0&V_e
 c_{13} s_{13} e^{-i\delta}\cr 0&0&0\cr
V_e c_{13} s_{13} e^{i\delta}&0&V_e s_{13}^2+
\frac{\Delta m_{atm}^2}{2E}&}+\frac{\Delta m^2_{sol}}{2E}
\pmatrix{s_{12}^2&s_{12}c_{12}&0\cr s_{12}c_{12}&c_{12}^2&0\cr0&0&1} 
\ea

From this we obtain $\nu'(x)=S'\nu'(0)$, with
\be
S'=e^{-iD L}
\pmatrix{\cos{\big(\frac{\omega ' L}{2}\big )}+
\frac{i}{\omega '} A'
\sin{\big(\frac{\omega 'L}{2}\big )}&0&-\frac{i}{\omega '}
B'e^{-i\delta}
\sin{\big (\frac{\omega ' L}{2}\big )}\cr
0&e^{iD L}&0\cr
-\frac{i}{\omega '}B'e^{i\delta}
\sin{\big (\frac{\omega 'L}{2}\big )}&0&
\cos{\big(\frac{\omega ' L}{2}\big )}-
\frac{i}{\omega '}A'
\sin{\big(\frac{\omega ' L}{2}\big )}}
+\frac{\Delta m^2_{sol}}{2E} S'_c
\ee

\begin{figure}                             
\begin{center}
\mbox{\epsfxsize=10truecm
\epsffile{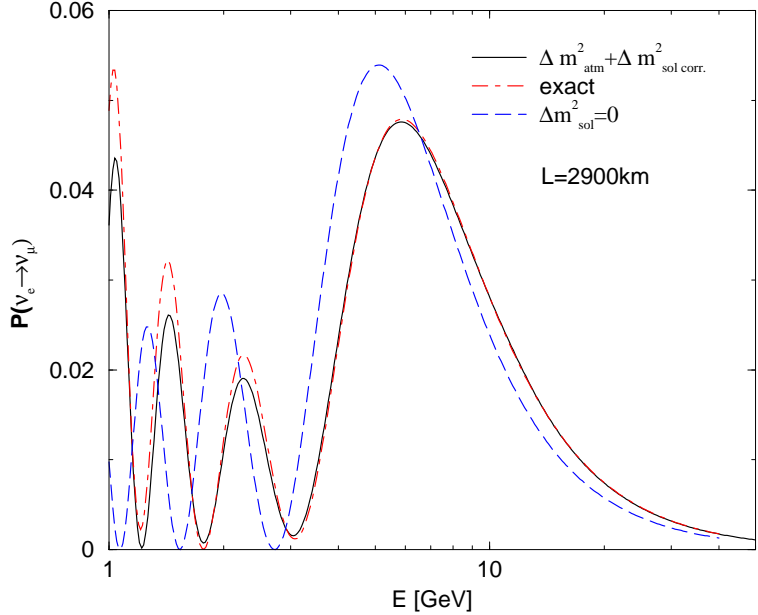}}
\end{center}
\caption{\footnotesize{
Plot of $P(\nu_e \to \nu_\mu)$ as a function of $E$ for $L=2900$ km.  Other
input parameter values are given in the text.}}
\label{fig:exp1}
\end{figure}

\begin{figure}                             
\begin{center}
\mbox{\epsfxsize=10truecm
\epsffile{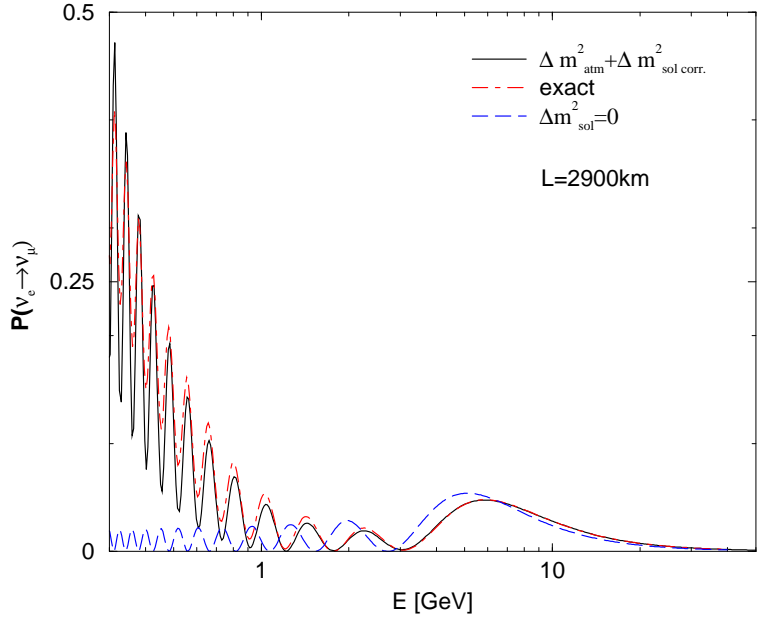}}
\end{center}
\caption{\footnotesize{Plot of $P(\nu_e \to \nu_\mu)$ as a function of $E$ for
$L=2900$ km, with detail of the region from $E=0$ to $E=1$ GeV.  Input
parameter values are given in the text.}}
\label{fig:exp2}
\end{figure}

The first term gives the one-$\Delta m^2$ contribution, which, after 
the rotation to the flavor basis $(\nu_e,\nu_\mu,\nu_\tau)$, is the
same as the one obtained in the previous section. The second term contains 
the corrections due to the $\Delta m^2_{sol}$ term. $S'_c$ is given 
by
\be
{S'_c}_{11}=e^{-iD L}\bigg[
L\frac{A'}{\omega'} s_{12}^2\sin{\big (\frac{\omega' L}{2}\big )}+
i\bigg (c_{12}^2 \frac{B'^2}{\omega'^3}
\sin{\big (\frac{\omega' L}{2}\big )}-
\frac{L}{2}\bigg(1+s_{12}^2 -c_{12}^2 \frac{A'^2}{2\omega'^2}\bigg )
\cos{\big (\frac{\omega' L}{2}\big ) }\bigg )
\bigg]
\ee
\be
{S'_c}_{22}=0
\ee
\be
{S'_c}_{33}=e^{-iD L}\bigg[
-L\frac{A'}{\omega'} \sin{\big (\frac{\omega' L}{2}\big )}-
i\bigg (c_{12}^2 \frac{B'^2}{\omega'^3}
\sin{\big (\frac{\omega' L}{2}\big )}+
L \frac{1}{2}\bigg(1+s_{12}^2 +c_{12}^2 \frac{A'^2}{2\omega'^2}\bigg )
\cos{\big (\frac{\omega' L}{2}\big ) }\bigg )
\bigg]
\ee
\be
{S'_c}_{12}=\frac{2E s_{12}c_{12}}{\Delta m^2_{atm}V_ec_{13}^2}
\bigg[
(V_ec_{13}^2-2D)
+
e^{-iD L} \bigg(
(2D-V_ec_{13}^2)
\cos{\big(\frac{\omega' L}{2}\big)}+ 
i\frac{\omega'^2+2A'D}{2\omega'}
\sin{\big(\frac{\omega' L}{2}\big)}\bigg) \bigg ] 
\ee
\be
{S'_c}_{13}=\frac{V_ec_{13}s_{13}e^{-i\delta}}{\omega'}e^{-iD L}
\bigg[ L (1+s_{12}^2)\sin{\big(\frac{\omega' L}{2}\big)}+ic_{12}^2\frac{ A'}
{\omega'} \bigg (\frac{2 }{\omega'}\sin{\big(\frac{\omega' L}{2}\big)}
-L\cos{\big(\frac{\omega' L}{2}\big)}\bigg)
\bigg]
\ee
\be
{S'_c}_{23}=\frac{2E}{\Delta m^2_{atm}}{s_{12} c_{12}}
{\rm tg}
\theta_{13}e^{-i\delta}
\bigg[1-e^{-iD L} \bigg(\cos{\big(\frac{\omega' L}
{2}\big)}+2i\frac{D}{\omega'} 
\sin{\big(\frac{\omega' L}{2}\big)}\bigg)
\bigg]
\ee
\be
{S'_c}_{21}={S'_c}_{12},\,\,\,\,{S'_c}_{31}=e^{2i\delta}{S'_c}_{13},
\,\,\,\,{S'_c}_{32}=e^{2i\delta}{S'_c}_{23}
\ee
with
\be
A'=\frac{\Delta m_{atm}^2}{2E}-{\sqrt{2}}G_FN_e\cos(2\theta_{13})
\ee
\be
B'=V_e\sin(2\theta_{13})
\ee
\be
\omega '=\sqrt{A'^2+B'^2}= \bigg [\bigg ( V_e\cos(2\theta_{13})-
\frac{\Delta m_{atm}^2}{2E}\bigg )^2+V_e^2\sin^2(2\theta_{13}) \bigg ]^{1/2}
\ee
After we rotate back in the actual flavor basis, the oscillation 
probabilities will be given by
\be
P(\nu_a\to\nu_b)=|(U R_{12}^\dagger S'R_{12} U^\dagger )_{ab}|^2
\label{pcorr}
\ee
The results from Eq. (\ref{pcorr}) are compared in Fig. \ref{fig:exp1},
\ref{fig:exp2} with those obtained from the exact numerical result and with the
result for the one-$\Delta m^2$ approximation for the following set of input
values: $\sin^2 2\theta_{23}=1$, $\Delta m^2_{32} = 3 \times 10^{-3}$ eV$^2$,
$\sin^2 2\theta_{sol}=\sin^2 2\theta_{12}=.8$ and $\Delta m^2_{sol}=\Delta
m^2_{21}=10^{-4}$eV$^2$ (or $\Delta m^2_{sol}=0$ for
comparison), and $\sin^2 2\theta_{13}=0.04$.

\begin{thebibliography}{99} 

\bibitem{sol}{Homestake: B.T. Cleveland et al, Astrophys.J. {\bf 496}, 505
(1998); Kamiokande: K. Hirata et al., Phys. Rev. {\bf D44}, 2241 (1991); SAGE:
J. N. Abdurashitov al., Phys. Rev. {\bf C60}, 055801 (1999); GALLEX: Hampel et
al., Phys. Lett. {\bf B447}, 127 (1999); GNO: M. Altmann et al., Phys.Lett.{\bf
B490}, 16 (2000); Superkamiokande: Y. Fukuda et al., Phys. Rev. Lett. {\bf 82},
1810, 243 (1999);  Phys. Rev. Lett. {\bf 86}, 5651, 5656 (2001); SNO: . 
Q. Ahmad et al., Phys. Rev. Lett. {\bf 87}, 071301 (2001).} 

\bibitem{fitsol}{In addition to the fits given by experimentalists themselves,
there are numerous fits by theorists; a few recent ones are J. Bahcall,
P. Krastev, and A. Smirnov, Phys. Rev. {\bf D60}, 093001 (1999); JHEP {\bf
0105}, 015 (2001); M. C. Gonzalez-Garcia, M. Maltoni, C. Pe\~na-Garay, and
J. W. F. Valle, Phys. Rev. {\bf D63}, 033005 (2001); J. N. Bahcall,
M. C. Gonzalez-Garcia, C. Pena-Garay, hep-ph/ hep-ph/0106258; P. Krastev,
A. Smirnov, hep-ph/0108177; G. L. Fogli, E. Lisi, A. Marrone, D. Montanino,
A. Palazzo, hep-ph/0104221; G. L. Fogli, E. Lisi, and A. Palazzo,
hep-ph/0105080; S. Choubey, S. Goswami, N. Gupta, D. P. Roy, hep-ph/0103318.}

\bibitem{othersol}{M.V. Garzelli, C. Giunti,hep-ph/0007155;
R. Barbieri, A. Strumia, JHEP {\bf 0012}, 016 (2000);
P. Creminelli, G. Signorelli, A. Strumia, JHEP 0105:052 (2001); 
V. Barger, D. Marfatia, K. Whisnant, B.P. Wood, Phys. Rev. {\bf D64} 073009
(2001).} 

\bibitem{msw}{L. Wolfenstein, Phys. Rev. {\bf D17}, 2369 (1978);
S. P. Mikheyev and A. Smirnov, Yad. Fiz. {\bf 42}, 1441 (1985)
[Sov.J. Nucl. Phys. {\bf 42}, 913 (1986)], Nuovo Cim., {\bf C9}, 17 (1986); 
S. P. Rosen and J. Gelb, Phys. Rev. {\bf D34}, 969 (1986); S. Parke,
Phys. Rev. Lett. {\bf 57}, 1275 (1986); W. Haxton, Phys. Rev. Lett. {\bf 57},
1271 (1986); J. Pantaleone and T. K. Kuo, Rev. Mod. Phys. {\bf 61}, 937
(1989).} 

\bibitem{kam}{K. S. Hirata et al., Phys. Lett. {\bf B205}, 416;
{\it ibid.}  {\bf 280}, 146 (1992); Y.Fukuda et al., Phys. Lett. {\bf B335},
237 (1994); S. Hatakeyama et al. Phys. Rev. Lett. {\bf 81}, 2016 (1998).} 

\bibitem{imb}{D. Casper et al., Phys. Rev. Lett. {\bf 66}, 2561
(1991); R.Becker-Szendy et al., Phys. Rev. {\bf D46}, 3720 (1992);
Phys. Rev. Lett.  {\bf 69}, 1010 (1992).} 

\bibitem{soudan2}{W. Allison et al., Phys. Lett. {\bf B391}, 491 (1997);
Phys. Lett. {\bf B449}, 137 (1999); W. A. Mann, Nucl. Phys. Proc. Suppl. {\bf
91}, 134 (2000).}

\bibitem{skatm}{Y. Fukuda {\it et al.}, Phys. Lett. {\bf B433}, 9 (1998);
Phys. Lett.\ {\bf B436}, 33 (1998).  Y. Fukuda {\it et al.}, Phys. Lett. {\bf
B467}, 185 (1999); Phys. Rev. Lett. {\bf 82}, 2644 (1999); C. McGrew, in the
Workshop on Neutrino Telescopes, Venice (Mar., 2001); C. Yanagisawa, in 
Frontiers in Contemporary Physics (Vanderbilt Univ., Mar. 2001); T. Toshito, 
in the Recontres de Moriond (Mar., 2001).} 

\bibitem{macro}{MACRO Collab., M. Ambrosio et al., Phys. Lett. {\bf B478}, 5
(2000); hep-ex/0001044; B. Barish, Nucl. Phys. Proc. Suppl. {\bf 91}, 141
(2000); hep-ex/0101019.}

\bibitem{chooz}{M. Apollonio et al., Phys. Lett. {\bf B420}, 397 (1998);
Phys. Lett. {\bf B466}, 415 (1999).}

\bibitem{paloverde}{F. Boehm et al., Phys. Rev. Lett. {\bf 84}, 3764 (2000); 
Phys. Rev.{\bf D62}, 072002 (2000).} 

\bibitem{k2k}{S. Ahn et al., Phys. Lett. {\bf B511}, 178 (2001); J. Hill, talk
at Snowmass-2001.} 

\bibitem{lsnd}{C. Athanassopoulous  et al., Phys. Rev. Lett.
{\bf 77}, 3082 (1996); Phys. Rev. Lett. {\bf 81}, 1774 (1998); A. Aguilar et
al., hep-ex/0104049.}
 
\bibitem{karmen}{K. Eitel (for KARMEN Collab.), Nucl. Phys. Proc. Suppl. 
{\bf 91}, 191 (2000).} 

\bibitem{lsndn}{If one were also to fit the LSND data, one would be led to
include light electroweak-singlet neutrinos. Since the LSND experiment has not
so far been confirmed, we shall not try to include the LSND data in our
discussion.} 

\bibitem{bgg}{A. Baltz, A. S. Goldhaber, M. Goldhaber, Phys. Rev. Lett. {\bf
81}, 5730 (1998).}

\bibitem{full}
G. L. Fogli, E. Lisi, A. Marrone, G. Scioscia, Phys. Rev. {\bf D59}, 033001 
(1998); C. W. Kim, U.W. Lee, Phys. Lett. {\bf B444}, 204 (1998); 
C. Giunti, , C. W. Kim, U. W. Lee, V.A. Naumov, hep-ph/9902261;
O. Peres, A. Yu. Smirnov, Phys. Lett. {\bf B456}, 204 (1999); 
A. Strumia, JHEP 9904,026 (1999).

\bibitem{foglidem}{G. L. Fogli, E. Lisi, and A. Marrone, Phys. Rev. {\bf D64},
093005 (2001).} 

\bibitem{bds}{G. Barenboim, A. Dighe, and S. Skadhauge, hep-ph/0106002.}

\bibitem{j}{C. Jarlskog, Z. Phys. {\bf C29}, 491 (1985); Phys. Rev. {\bf D35},
1685) (1987).} 

\bibitem{darkside}{A. de Gouvea, A. Friedland and H. Murayama, Phys. Lett.
{\bf B490}, 125 (2000).} 

\bibitem{kamland}{See, e.g., J. Busenitz et al., Proposal for
U.S. Participation in KamLAND (Mar. 1999). In addition to the reactor
$\bar\nu_e$ oscillation search, this experiment also intends to measure solar
$^7$Be neutrinos.} 

\bibitem{bk}{V. Barger, D. Marfatia, and B. Wood,
Phys. Lett. {\bf B498}, 53 (2001); H. Murayama, hep-ph/0012075.}

\bibitem{minos}{See http://www.hep.anl.gov/ndk/hypertext/minos.html.} 

\bibitem{cngs}{See http://proj-cngs.web.cern.ch/proj-cngs/.} 

\bibitem{jhf}{For information on the JHF-SuperK project and possible upgrades
involving a more intense beam and the construction of a 1 Mton water Cherenkov
far detector (HyperKamiokande), see http://neutrino.kek.jp/jhfnu, in
particular, Y. Itoh et al., ``Letter of Intent: A Long Baseline Neutrino
Oscillation Experiment using the JHF 50 GeV Proton Synchrotron and the
Super-Kamiokande Detector'' (Feb. 2000); Y. Itoh et al, ``The JHF-Kamioka
Neutrino Project''.} 

\bibitem{superbeam}{V. Barger et al. (Fermilab Superbeam Working Group), Nov.,
2000, hep-ph/0103052; Phys. Rev. {\bf D63}, 113011 (2001); D. Harris, talk at
UNO workshop (Aug. 2000), R. Shrock, talks at UNO workshop, June, 2001;
http://superk.physics.sunysb.edu/uno; J. Cadenas et al. (CERN Superbeam 
Working Group), hep-ph/0105297.  Some possibilities include a second
detector (perhaps off-axis) as an extension of the MINOS experiment, and
options involving FNAL-Homestake, BNL-Ithaca, BNL-Homestake, CERN-Frejus, or 
JHF-Beijing.} 

\bibitem{anl}{References and websites for these experiments and future projects
can be found, e.g., at http://www.hep.anl.gov/ndk/hypertext/nu\_industry.html.}

\bibitem{geer}{S. Geer, Phys. Rev. {\bf D57}, 6989 (1998).} 

\bibitem{fnal}{C. Albright et al. (Fermilab Neutrino Factory Physics Working
Group), ``Physics at a Neutrino Factory'', Apr. 2000, hep-ph/hep-ex/0008064. 
See also the companion machine study ``Feasibility Study of a Neutrino Factory
Based on a Muon Storage Ring'', by the Fermilab Neutrino Factory Feasibility
Study Working Group, N. Holtkamp, et al.}

\bibitem{bnl}{Muon Collider and Neutrino Factory Collaboration, ``Neutrino
Factory Feasibility Study II'' (May 2001), available at
http://www.cap.bnl.gov/mumu.}

\bibitem{nufact}{See, e.g., 

http://www.fnal.gov/projects/muon\_collider/nu/study/study.html

http://www.cern.ch/\~\ autin/nufact99/whitepap.ps

http://www.cap.bnl.gov/mumu

http://puhep1.princeton.edu/mumu/NSFLetter/nsfmain.ps.} 

\bibitem{nufactmeasurements}

V. Barger, K. Whisnant, S. Pakvasa, and R. J. N. Phillips, Phys. Rev. 
{\bf D22}, 2718 (1980);
P. Krastev, Nuovo Cimento {\bf 103A}, 361 (1990);
R. H. Bernstein and S. J. Parke, Phys. Rev. {\bf D44}, 2069 (1991);
De Rujula, M. B. Gavela, and P. Hernandez, Nucl. Phys. 
{\bf B547}, 21 (1999);
P. Lipari, Phys.Rev. {\bf D61}, 113004 (2000) 
S. Dutta, R. Gandhi, and B. Mukhopadhyaya, Eur.Phys.J. {\bf C18}, 405 (2000);
V. Barger, S. Geer, K. Whisnant, Phys.Rev. {\bf D61}, 053004 (2000); 
D. Dooling, C. Giunti, K. Kang, C. W. Kim,Phys.Rev. {\bf D61},073011 (2000);
A. Bueno, M. Campanelli, A. Rubbia, Nucl.Phys. {\bf B573}, 27 (2000); 
V. Barger, S. Geer, R. Raja, K. Whisnant, Phys.Rev. {\bf D62}, 013004 (2000); 
M. Freund, M. Linder, S.T. Petcov, A. Romanino, Nucl.Phys. {\bf B578}, 27
(2000);  
I. Mocioiu and R. Shrock, AIP Conf.Proc. {\bf 533}, 74 (2000) (NNN99); 
I. Mocioiu and R. Shrock, Phys.Rev. {\bf D62}, 053017 (2000);
V. Barger, S. Geer, R. Raja, K. Whisnant, Phys.Rev. {\bf D62}, 073002 (2000); 
A. Cervera, A. Donini, M.B. Gavela, J. Gomez Cadenas, P. Hernandez, O. Mena,
and S. Rigolin, Nucl.Phys. {\bf B579}, 17 (2000), Erratum-ibid. {\bf B593},
731 (2001); 
M. Freund, P. Huber, M. Lindner, Nucl.Phys. {\bf B585}, 105 (2000); 
V. Barger, S. Geer, R. Raja, K. Whisnant, Phys.Lett. {\bf B485}, 379 (2000); 
Z.-Z. Xing, Phys. Lett. {\bf487}, 327 (2000); Phys. Rev. {\bf D63}, 073012
(2000);  
A. Bueno, M. Campanelli, A. Rubbia, Nucl.Phys. {\bf B589}, 577 (2000); 
P. Fishbane, Phys. Rev. {\bf D62}, 093009 (2000); P. Fishbane and P. Kaus,
Phys. Lett. {\bf B506}, 275 (2001); P. Fishbane and S. Gasiorowicz, 
hep-ph/0012230; 
V. Barger, S. Geer, R. Raja, K. Whisnant, Phys.Rev. {\bf D63}, 033002 (2001); 
M. Freund, P. Huber, M. Lindner, hep-ph/0105071.
 
\bibitem{akh}
E. Akhmedov, A. Dighe, P. Lipari, A. Smirnov, Nucl. Phys. {\bf B542}, 3 (1999);
E. Akhmedov, Nucl.Phys. {\bf B538}, 25 (1999); E. Akhmedov, hep-ph/0001264;
E. Akhmedov, P. Huber, M. Lindner, and T. Ohlsson, 
Nucl. Phys. {\bf B608}, 394 (2001).

\bibitem{prem}
A.Dziewonski, Earth Structure, in: ''The Encyclopedia of Solid
Earth Geophysics'', D.E.James (Ed.), (Van Nostrand Reinhold, New York, 1989),
p.331.

\bibitem{exact}

See, e.g., H. Zaglauer, K. Schwarzer, Z. Phys. {\bf C40}, 273 (1988); 
T. Ohlsson, H. Snellman,  J. Math. Phys. {\bf 41}, 2768 (2000).

\bibitem{perturbation}
 J. Arafune, J. Sato, Phys.Rev. {\bf D55}, 1653 (1997); 
H. Minakata, H. Nunokawa, Phys.Rev. {\bf D57}, 4403 (1998); 
M. Koike, J. Sato, Phys. Rev. {\bf D62}, 073006 (2000);
 
\bibitem{cp}
S.M. Bilenky, C. Giunti, W.Grimus, Phys.Rev. {\bf D58}, 033001 (1998); 
K. Dick, M. Freund, M. Lindner,  A. Romanino, Nucl. Phys. {\bf B562}, 29 
(1999); 
M. Tanimoto, Phys. Lett. {\bf B462}, 115 (1999); 
A. Donini, M.B. Gavela, P. Hernandez, S. Rigolin, Nucl.Phys. {\bf B574}, 23 
(2000);
A. Romanino, Nucl.Phys. {\bf B574} 675 (2000);
 M. Koike, J. Sato, Phys.Rev. {\bf D61}, 073012 (2000), Erratum-ibid. 
{\bf D62}, 079903 (2000);
S. J. Parke, T. J. Weiler, Phys.Lett. {\bf B501}, 106 (2001);
T. Miura, E. Takasugi, Y. Kuno, and M. Yoshimura, Phys. Rev. {\bf D64}, 013002
(2001); M. Koike, T. Ota, J. Sato, hep-ph/0011387; P. Lipari. hep-ph/0102046.

\end{thebibliography}
\end{document}